\documentclass[a4paper]{article}
\pdfoutput=1
\usepackage[utf8]{inputenc}
\usepackage[T1]{fontenc}
\usepackage{lmodern, amsmath, a4wide, xcolor, endnotes}
\usepackage[numbers, elide, sort&compress]{natbib}

\makeatletter\g@addto@macro\bfseries{\boldmath}\makeatother

\usepackage[colorlinks=true, citecolor=blue, urlcolor=blue, linkcolor=blue, breaklinks=true, pdfpagelabels=false]{hyperref}
\def\figureautorefname~#1\null{figure\,#1\null}

\def\equationautorefname~#1\null{equation\,(#1)\null}
\newcommand{\fullref}[2]{\hyperref[#2]{#1\,\ref*{#2}}}
\makeatletter
\newcommand{\rcite}[1]{\hyper@@link[cite]{}{cite.#1}{ref.\,\cite*{#1}}}
\makeatother
\newcommand*{\eqsref}[2]{\hyperref[#1]{Eqs.\,(\ref*{#1}--}\hyperref[#2]{\ref*{#2})}}

\let\oldhref\href
\renewcommand{\href}[2]{%
	\oldhref{#1}{#2}%
	\endnote{\url{#1}}%
}

\allowdisplaybreaks[4]

\let\Re\undefined
\DeclareMathOperator{\Re}{Re}

\newcounter{affiliation}
\newcommand{\aff}[1]{$^{\ref{aff:#1}}$ }
\newcommand{\twoaff}[2]{$^{\ref{aff:#1},\ref{aff:#2}}$ }

\newcommand{\naff}[2]{\noindent\refstepcounter{affiliation}\label{aff:#1}$^{\arabic{affiliation}}$\,#2\\}

\title{\vspace*{-37mm}{\normalsize\hfill CERN-LPCC-2019-02\\}\vspace*{7mm}%
Proposal for the validation of Monte Carlo implementations\\of the standard model effective field theory}

\author{\hspace*{-0.05\textwidth}\parbox{1.1\textwidth}{\centering
Gauthier Durieux\aff{Technion} (ed.),
Ilaria Brivio\twoaff{NBIA}{Heidelberg} (ed.),\\
Fabio Maltoni\twoaff{UCLouvain}{Bologna}  (ed.\ ex officio),
Michael Trott\aff{NBIA}  (ed.\ ex officio),\\
Simone Alioli,\aff{Bicocca-INFN}
Andy Buckley,\aff{Glasgow}
Mauro Chiesa,\aff{Wuerzburg}
Jorge de Blas,\twoaff{Padova}{INFN-Padova}
Athanasios Dedes,\aff{ioamnina}
Céline Degrande,\aff{UCLouvain}
Ansgar Denner,\aff{Wuerzburg}
Christoph Englert,\aff{Glasgow}
James Ferrando,\aff{DESY}
Benjamin Fuks,\twoaff{LPTHE}{IUF}
Peter Galler,\aff{Glasgow}
Admir Greljo,\aff{CERNTH}
Valentin Hirschi,\aff{ETH}
Gino Isidori,\aff{Zurich}
Wolfgang Kilian,\aff{Siegen}
Frank Krauss,\aff{Durham}
Jean-Nicolas Lang,\aff{Zurich}
Jonas Lindert,\aff{Durham}
Michelangelo Mangano,\aff{CERNTH}
David Marzocca,\aff{Trieste-INFN}
Olivier Mattelaer,\aff{UCLouvain}
Kentarou Mawatari,\aff{Osaka}
Emanuele Mereghetti,\aff{LosAlamos}
David J.\ Miller,\aff{Glasgow}
Ken Mimasu,\aff{UCLouvain}
Michael Paraskevas,\aff{Warsaw}
Tilman Plehn,\aff{Heidelberg}
Laura Reina,\aff{fsu}
Janusz Rosiek,\aff{Warsaw}
Jürgen Reuter,\aff{DESY}
José Santiago,\aff{Granada}
Kristaq Suxho,\aff{ioamnina}
Lampros Trifyllis,\aff{ioamnina}
Eleni Vryonidou,\aff{CERNTH}
Christopher White,\aff{QueenMary}
Cen Zhang,\twoaff{IHEP}{Beijing}
Hantian~Zhang\aff{Zurich}
\\[5mm]\footnotesize
\naff{Technion}{Physics Department, Technion-Israel Institute of Technology, Haifa 32000, Israel}
\naff{NBIA}{Niels Bohr International Academy and Discovery Center, Niels Bohr Institute, University of Copenhagen, 2100 Copenhagen, Denmark}
\naff{Heidelberg}{Institut für Theoretische Physik Universität Heidelberg, 69120 Heidelberg, Germany}
\naff{UCLouvain}{Centre for Cosmology, Particle Physics and Phenomenology (CP3), Université catholique de Louvain, 1348 Louvain-la-Neuve, Belgium}
\naff{Bologna}{Dipartimento di Fisica e Astronomia, Università di Bologna and INFN, Sezione di Bologna, via Irnerio 46, 40126 Bologna, Italy}
\naff{Bicocca-INFN}{Universit\`a degli Studi di Milano-Bicocca and INFN, Sezione di Milano-Bicocca, 20126, Milano, Italy}
\naff{Glasgow}{School of Physics and Astronomy, University of Glasgow, Glasgow, G12 8QQ, UK}
\naff{Wuerzburg}{Julius-Maximilians-Universität Würzburg, Institut für Theoretische Physik und Astrophysik, Emil-Hilb-Weg 22, 97074 Würzburg, Germany}
\naff{Padova}{Dipartimento di Fisica e Astronomia “Galileo Galilei”, Università di Padova, Via Marzolo 8, 35131 Padova, Italy}
\naff{INFN-Padova}{INFN, Sezione di Padova, Via Marzolo 8, 35131 Padova, Italy}
\naff{ioamnina}{Department of Physics, Division of Theoretical Physics, University of Ioannina, 45110, Greece}
\naff{DESY}{DESY, Notkestraße 85, 22607 Hamburg, Germany}
\naff{LPTHE}{Laboratoire de Physique Th\'eorique et Hautes Energies (LPTHE), UMR 7589, Sorbonne Universit\'e et CNRS, 4 place Jussieu, 75252 Paris Cedex 05, France}
\naff{IUF}{Institut Universitaire de France, 103 boulevard Saint-Michel, 75005 Paris, France}
\naff{CERNTH}{CERN, TH Department, Geneva, Switzerland}
\naff{ETH}{ETH Zürich, Rämistrasse 101, 8092 Zürich, Switzerland}
\naff{Zurich}{Physik-Institut, Universität Zürich, 8057 Zürich, Switzerland}
\naff{Siegen}{University of Siegen, Department of Physics, 57068 Siegen, Germany}
\naff{Durham}{Institute for Particle Physics Phenomenology, Department of Physics, Durham University, Durham DH1 3LE, UK}
\naff{Trieste-INFN}{INFN - Sezione di Trieste, Via Bonomea 265, 34136, Trieste, Italy}
\naff{Osaka}{Department of Physics, Osaka University, Toyonaka, Osaka 560-0043, Japan}
\naff{LosAlamos}{Theoretical Division, Los Alamos National Laboratory, Los Alamos, NM 87545, USA}
\naff{Warsaw}{Faculty of Physics, University of Warsaw, Pasteura 5, 02-093 Warsaw, Poland}
\naff{fsu}{Department of Physics, Florida State University, 510 Keen Building, Tallahassee, FL 32306-4350}
\naff{Granada}{CAFPE and Departamento de Física Teórica y del Cosmos, Universidad de Granada, 18071 Granada, Spain}
\naff{RomeI-INFN}{INFN, Sezione di Roma, Piazzale Aldo Moro 2, 00185 Roma, Italy}
\naff{QueenMary}{Centre for Research in String Theory, School of Physics and Astronomy, Queen Mary University of London, 327 Mile End Road, London E1 4NS, UK}
\naff{IHEP}{Institute of High Energy Physics, Chinese Academy of Sciences, Beijing 100049, China}
\naff{Beijing}{School of Physical Sciences, University of Chinese Academy of Sciences, Beijing 100049, China}
}}
\date{}

\begin{document}

\thispagestyle{empty}
\maketitle
\renewcommand{\abstractname}{\vspace{-2mm}Abstract}
\begin{abstract}
We propose a procedure to cross-validate Monte Carlo implementations of the standard model effective field theory. It is based on the numerical comparison of squared amplitudes computed at specific phase-space and parameter points in pairs of implementations. Interactions are fully linearised in the effective field theory expansion. The squares of linear effective field theory amplitudes and their interference with standard-model contributions are compared separately. Such pairwise comparisons are primarily performed at tree level and a possible extension to the one-loop level is also briefly considered. We list the current standard model effective field theory implementations and the comparisons performed to date.
\end{abstract}


\section{Introduction}

This note proposes a procedure to validate Monte Carlo (MC) implementations ---more or less partial--- of the standard model effective field theory (SMEFT).
The underlying agreement was achieved under the auspices of the LHC Top and Electroweak Working Groups, and of the LHC Higgs Cross Section Working Group. 

Most of the existing MC implementations are able to compute leading-order predictions. Therefore, we primarily consider tree-level calculations. Many implementations are based on Lagrangians written in codes such as \textsc{Sarah}~\cite{Staub:2015kfa} or \textsc{FeynRules}~\cite{Alloul:2013bka} and then exported through standard format to matrix-element-based MC generators. A systematic procedure for comparing MC implementations is outlined through the comparison of squared amplitudes.

\section{Existing implementations}
A brief list of public SMEFT implementations with links and references is provided below.
It shall be updated as new implementations become available.
\begin{itemize}
\item \textsc{dim6top} is a UFO implementation of top-quark interactions following the conventions of the LHC top Working Group~\cite{AguilarSaavedra:2018nen}. It is available at this \href{https://feynrules.irmp.ucl.ac.be/wiki/dim6top}{url}.

\item \textsc{SMEFTsim} is a complete UFO implementation of the Warsaw basis~\cite{Grzadkowski:2010es} of dimension-six operators~\cite{Brivio:2017btx}. It is available at this \href{https://feynrules.irmp.ucl.ac.be/wiki/SMEFT}{url}.

\item \textsc{SMEFT@NLO} is a UFO implementation, to next-to-leading order in QCD, of of CP- and $U(2)_q\times U(2)_u\times U(3)_d\times U(3)_l\times U(3)_e$-conserving dimension-six interactions, available at this \href{https://feynrules.irmp.ucl.ac.be/wiki/SMEFTatNLO}{url}.

\item \textsc{SmeftFR}~\cite{Dedes:2017zog,Dedes:2019uzs} is a package generating Feynman rules, in \textsc{FeynRules} and UFO formats, for the dimension-six operators of the Warsaw basis~\cite{Grzadkowski:2010es} (or any subset), in unitary or linear $R_\xi$ gauges, in terms of physical fields (mass eigenstates), for general flavour structures. It is available at this \href{http://www.fuw.edu.pl/smeft}{url}.

\item \textsc{HEL}~\cite{Alloul:2013naa} is an implementation of dimension-six operators in the SILH basis~\cite{Giudice:2007fh} available at this \href{https://feynrules.irmp.ucl.ac.be/wiki/HEL}{url}.

\item \textsc{BSMC}~\cite{Falkowski:2015wza} is an implementation of dimension-six operators in the Higgs basis~\cite{Falkowski:2001958} associated with the \textsc{Rosetta} package (\href{https://rosetta.hepforge.org}{here}). It is available at this \href{http://feynrules.irmp.ucl.ac.be/wiki/BSMCharacterisation}{url}.

\item \textsc{Powheg SM-EFT}~\cite{Alioli:2018ljm} is an implementation of SMEFT corrections to neutral and charged current Drell-Yan and electroweak Higgs boson production, including next-to-leading order QCD corrections and interfaced with parton showering and hadronization according to the POWHEG method. It is available at this \href{http://powhegbox.mib.infn.it/}{url}.

\item \textsc{Recola2} is a  tree-level and one-loop matrix-element provider. While \textsc{Recola}~\cite{Actis:2012qn,Actis:2016mpe} was specifically designed for the SM, \textsc{Recola2}~\cite{Denner:2017vms,Denner:2017wsf} can be linked against any 
user-provided model file. For the present comparison, a tree-level model containing the 
full set of dimension-six operators in the Warsaw basis was implemented. 
A preliminary version of the model was used in ref.~\cite{Chiesa:2018lcs} 
to study the impact of the dimension-six operators contributing to anomalous
triple-gauge-boson interactions in vector-boson pair production at the LHC.

\item \textsc{SFitter} is a general purpose fitting tool that combines a large number of experimental channels in a global fit to constrain dimension-six operators in the HISZ basis.
The basis is implemented in a UFO file and processed by \textsc{MadGraph5}, yielding parametrizations of experimental observables in terms of the Wilson coefficients.
This implementation, extensively validated internally, is not public.

\item \textsc{HiggsPO} is a UFO implementation of the Higgs pseudo-observables framework at next-to-leading order in QCD~\cite{Gonzalez-Alonso:2014eva,Greljo:2015sla,Greljo:2017spw}, available at this \href{https://www.physik.uzh.ch/data/HiggsPO}{url}.

\end{itemize}

\section{Validation procedure}
\label{sec:validation}

The cross-validation of MC implementations of new physics models is in general a very time-consuming exercise, especially in the case of EFTs which feature large number of new interactions and parameters.
In addition, as existing implementations rely on different operator bases, conventions and tools, a flexible, robust and scalable approach needs to be employed.
The main idea of our proposal is to use physical predictions as terms of comparison.
These are by definition independent of a certain number of conventions, and can be uniquely defined once the input parameters, the scheme employed and the order of the computations are set. 

A first possibility is to compare total cross sections for producing given final states. Cross sections have the advantage of being easily obtainable in any MC framework. Such procedure was for example applied in \rcite{AguilarSaavedra:2018nen} for cross-validating the implementations of the top-quark sectors in \texttt{SMEFTsim} and \texttt{dim6top} frameworks.
However, total cross sections are intrinsically limited by the numerical precision of the phase-space integration, which in some cases can entail considerable computing time.
For example, to test operators involving four top quarks, processes such as $pp\to t\bar{t}t\bar{t}$ have to be considered and corresponding cross sections computed.
We therefore adopt here a much leaner strategy that has been employed successfully in the past to compare models in \textsc{FeynRules}~\cite{Alloul:2013bka} with native ones in generators such as \textsc{MadGraph5}~\cite{Alwall:2014hca}, \textsc{Sherpa}~\cite{Gleisberg:2008ta}, \textsc{Whizard}~\cite{Kilian:2007gr}, \textsc{Herwig++}~\cite{Bahr:2008pv}, \textsc{CalcHEP}~\cite{Belyaev:2012qa} and \textsc{FeynArts/FormCalc}~\cite{Hahn:2000kx, Hahn:2016ebn}.

We propose the following validation procedure:
\begin{enumerate}
\item Comparisons are performed between pairs of models, at a series of parameter points.

One may not aim at a fully connected network of $n$ model with $n^2$ comparisons, a sparser one may be sufficient. It is expected that families of similar implementations that are easier to compare among themselves will form.

\item SM and EFT predictions are primarily considered at the tree level (see \autoref{sec:loop} for an extension to the one-loop level).

\item The same set of electroweak input parameters is to be used in both models when comparing operators that shift them.
Masses ($m_Z$ and $m_W$) are included as input parameters to avoid SMEFT corrections to propagators that are not easy to treat in MC programs.

Standard numerical values are input parameters for instance $\{G_F=1.16637\cdot10^{-5}\,\text{GeV}^{-2},\linebreak m_Z=91.1876\,\text{GeV}, m_W=80.379\,\text{GeV},m_h=125\,\text{GeV}, m_t=172\,\text{GeV}, \alpha_S(m_Z)=0.1184\}$.

We also suggest to fix the CKM matrix to unity, unless another scheme is adopted between the two models compared.

\item Particle widths are explicitly set to zero in the comparison, to avoid differences due to gauge choices in the two models.
SMEFT corrections to propagators, that are not easy to treat for MC programs, would otherwise again be generated.

The dependence of decay amplitude squared on EFT parameters could also be explicitly examined through one-to-$n$ squared amplitudes. One may also adopt the complex mass scheme in the future.

\item \label{item:linearisation} The couplings of EFT vertices are linearised in the $C/\Lambda^2$ expansion.
A linearisation of all internal parameters (such as coupling constants, the Higgs vev, electroweak inputs, etc.) and of their functions (denominators, square roots, power, etc.) is in particular required.
This allows for the isolation by MC programs of amplitudes themselves linear in this expansion. Equivalent parameter points in different implementations are then also obtained through a linear mapping.

\item \label{item:squared-amplitudes} The computation of a set of two-to-$n$ (and one-to-two) squared amplitudes, averaged over initial colours and helicities, summed over final ones, is performed at given phase space points (which may not necessarily physically representative).
Numerical values are quoted in units of GeV to the appropriate power.

In the future, one could consider comparing helicity amplitudes squared.

When targetting Higgs or top-quark EFT interactions, the set of all SM two-to-two  processes (excluding crossing-symmetric and conjugates) involving at least one of these particles can for instance be considered for comparisons, together with selected two-to-three processes like \verb;g g > t t~ h; and \verb;q q~ > q q~ h/a/z/w; processes (see \autoref{tab:processes}).

\item In addition to the SM contribution $|A_\text{SM}|^2$, the interference $2\Re\{A_\text{SM}^*A_\text{EFT}\}$ and squared $|A_\text{EFT}|^2$ EFT contributions are evaluated for equivalent parameter points in the two models.

Although the $|A_\text{EFT}|^2$ contributions do not constitute complete EFT results to $1/\Lambda^4$ order, they are unambiguously defined (since they arise from the square of leading EFT contributions at the amplitude level which are themselves unambiguously defined) and useful for the comparison of implementations, in particular in cases where the $2\Re\{A_\text{SM}^*A_\text{EFT}\}$ interference vanishes.

Considered parameter points may for instance correspond to one single operator coefficient set to a non-vanishing value in one of the two models compared.
One should then remember to check the relative sign of operators having vanishing interferences with the SM model by considering parameter points where two of them are non vanishing.
Establishing sets of equivalent parameter points in implementations relying on different operator bases may require some effort.

\item Information about the implementations compared, parameter points, phase-space points, and averaged squared amplitudes are stored in LHE format (see v1~\cite{Alwall:2006yp}, \href{http://www.lpthe.jussieu.fr/LesHouches09Wiki/index.php/LHEF_for_Matching}{v2}~\cite{Butterworth:2010ym} and \href{https://phystev.cnrs.fr/wiki/2013:groups:tools:lhef3}{v3} descriptions).

Information about the generators, model versions, processes and parameter points used to compute the amplitude squared corresponding to a given \texttt{id} are provided under the standard \texttt{<weight>} header (see \autoref{tab:lhe} for an example).
The generator version used is specified in a \texttt{<weight\_generator>} sub-header, the model version in a \texttt{<weight\_model\_version>} one. The date at which a UFO model has been generated can for instance be specified here to be as specific as possible.
Information about the generated processes (numbered), the coupling order restrictions, etc.\ are specified in a generator-dependent format in a \texttt{<weight\_card>} sub-header.
The parameter point is provided in SLHA format in a \texttt{<weight\_slha>} sub-header.

The \texttt{<event>} field is used to store the phase-space point at which squared amplitudes are evaluated. The values of the latter for every \texttt{id} specified earlier are stored in a \texttt{<wgt>} field.
The \verb;id; label can be chosen to be human-readable and to include, separated by hyphens, information about the implementation, the parameter point, and a \verb;sq; or \verb;int; label specifying whether it corresponds to a squared or interference term  (see again \autoref{tab:lhe} for an example).

\begin{table}%
\begin{scriptsize}\begin{verbatim}
<LesHouchesEvents version="3.0">
  <header>
    ...
    <initrwgt>
      <weightgroup name='mg_reweighting' weight_name_strategy='includeIdInWeightName'>
        <weight id='mod1-sm-sq'>
          <weight_generator>MadGraph5_aMC@NLO_v2.6.5</weight_generator>
          <weight_model-version>dim6top_LO_UFO v1.0 Sat 8 Dec 2018 14:17:45</weight_model_version>
          <weight_card>
            change model SMEFTsim_A_general_MwScheme_UFO_v2-sm
            change process  g g > t~ t FCNC=0 QED<=99 QCD<=99 NP=0 @1
            change process  g a > t~ t FCNC=0 QED<=99 QCD<=99 NP=0 @2
            change process  g t > z  t FCNC=0 QED<=99 QCD<=99 NP=0 @3
            ...
          </weight_card>
          <weight_slha>
            Block EFT
               22 0. # ctB
               23 0. # ctW
            ...
          </weight_slha>
        </weight>
        <weight id='mod1-ctW-int'>
          <weight_generator>MadGraph5_aMC@NLO_v2.6.5</weight_generator>
          <weight_model_version>dim6top_LO_UFO v1.0 Sat 8 Dec 2018 14:17:45</weight_model_version>
          <weight_card>
            change model SMEFTsim_A_general_MwScheme_UFO_v2-ctW
            change process  g g > t~ t FCNC=0 QED<=99 QCD<=99 NP<=1 NP^2==1 @1
            change process  g a > t~ t FCNC=0 QED<=99 QCD<=99 NP<=1 NP^2==1 @2
            change process  g t > z  t FCNC=0 QED<=99 QCD<=99 NP<=1 NP^2==1 @3
            ...
          </weight_card>
          <weight_slha>
            Block EFT
               22 0. # ctB
               23 1. # ctW
            ...
          </weight_slha>
        </weight>
        ...
      </weightgroup>
    </initrwgt>
  </header>
  ...
  <event>
     4      2 +5.7925728e-01 9.11180000e+01 0.00000000e+00 1.18400000e-01
           21 -1    0    0    0    0 +0.00e+00 +0.00e+00 +5.00e+02 5.00e+02 0.00e+00 0.00e+00 9.00e+00
           22 -1    0    0    0    0 +0.00e+00 +0.00e+00 -5.00e+02 5.00e+02 0.00e+00 0.00e+00 9.00e+00
           -6  1    1    2    0    0 +1.66e+02 -3.90e+02 +2.01e+02 5.00e+02 1.72e+02 0.00e+00 9.00e+00
            6  1    1    2    0    0 -1.66e+02 +3.90e+02 -2.01e+02 5.00e+02 1.72e+02 0.00e+00 9.00e+00
    <rwgt>
    <wgt id='mod1-sm-sq'> +2.1166989e-01 </wgt>
    <wgt id='mod1-ctW-int'> +3.6769585e-01 </wgt>
    <wgt id='mod1-ctZ-int'> -3.2187585e-01 </wgt>
    <wgt id='mod1-ctG-int'> +3.0451935e-02 </wgt>
    <wgt id='mod1-ctW-sq'> +1.5438777e+00 </wgt>
    <wgt id='mod1-ctZ-sq'> +1.1830749e+00 </wgt>
    <wgt id='mod1-ctG-sq'> +1.0589251e-02 </wgt>
    <wgt id='mod2-sm-sq'> +2.1166988e-01 </wgt>
    <wgt id='mod2-ctW-int'> +3.6769591e-01 </wgt>
    <wgt id='mod2-ctZ-int'> -3.2187587e-01 </wgt>
    <wgt id='mod2-ctG-int'> +3.0451934e-02 </wgt>
    <wgt id='mod2-ctW-sq'> +1.5438783e+00 </wgt>
    <wgt id='mod2-ctZ-sq'> +1.1830750e+00 </wgt>
    <wgt id='mod2-ctG-sq'> +1.0589251e-02 </wgt>
    ...
    </rwgt>
  </event>
  ...
</LesHouchesEvents>
\end{verbatim}\end{scriptsize}%
\caption{Example of LHE format (based on the SLHA3 convention) including headers specifying the EFT implementations, processes (numbered with \texttt{@i} in \textsc{MadGraph} syntax), parameter points. The event record relative to a given process and parameter point specifies the phase-space point and the averaged squared amplitudes obtained for the specified implementations, parameter points and EFT order.}
\label{tab:lhe}
\label{tab:event}
\end{table}

\item Cross-validations realized by model authors will be referenced in \autoref{sec:comparisons} of this note. Existing LHE files can serve as basis of comparison for additional models and be complemented with the numerical values they give rise to.

\item Comparison authors are encouraged to implement the basis translation(s) established in the public tools \textsc{Rosetta}~\cite{Falkowski:2015wza} (available at this \href{https://rosetta.hepforge.org}{url}) and \textsc{wcxf-python}~\cite{Aebischer:2017ugx} (available at this \href{https://wcxf.github.io/python.html}{url}). Such tools can then be used to generate SLHA inputs for equivalent parameter points. Note that basis translations should be kept linear in the EFT expansion to comply with \autoref{item:linearisation}.
\end{enumerate}

\begin{table}\centering
\begin{scriptsize}\tt
\begin{tabular}[t]{@{\hskip-5mm}l}
\hfil Higgs\\\verb;g g > g h;\\\verb;g g > h h;\\\verb;g u > u h;\\\verb;g c > c h;\\\verb;g d > d h;\\\verb;g s > s h;\\\verb;g b > b h;\\\verb;g t > t h;\\\verb;u u~ > a h;\\\verb;u u~ > z h;\\\verb;u d~ > w+ h;\\\verb;c c~ > a h;\\\verb;c c~ > z h;\\\verb;c s~ > w+ h;\\\verb;d u~ > w- h;\\\verb;d d~ > a h;\\\verb;d d~ > z h;\\\verb;s c~ > w- h;\\\verb;s s~ > a h;\\\verb;s s~ > z h;\\\verb;b b~ > a h;\\\verb;b b~ > z h;\\\verb;b t~ > w- h;\\\verb;b~ t > w+ h;\\\verb;a a > h h;\\\verb;a ve > ve h;\\\verb;a vm > vm h;\\\verb;a vt > vt h;\\\verb;a e- > e- h;\\\verb;a mu- > mu- h;\\\verb;a ta- > ta- h;\\\verb;a t > t h;\\\verb;a z > h h;\\\verb;a w+ > w+ h;\\\verb;ve ve~ > z h;\\\verb;ve e+ > w+ h;\\\verb;vm vm~ > z h;\\\verb;vm mu+ > w+ h;\\\verb;vt vt~ > z h;\\\verb;vt ta+ > w+ h;\\\verb;e- ve~ > w- h;\\\verb;e- e+ > z h;\\\verb;mu- vm~ > w- h;\\\verb;mu- mu+ > z h;\\\verb;ta- vt~ > w- h;\\\verb;ta- ta+ > z h;\\\verb;t t~ > z h;\\\verb;t t~ > h h;\\\verb;z z > h h;\\\verb;z w+ > w+ h;\\\verb;w+ h > w+ h;\\\verb;h h > h h;
\end{tabular}
\begin{tabular}[t]{l}
\hfil top\\\verb;g g > t~ t;\\\verb;g b > w- t;\\\verb;g a > t~ t;\\\verb;g t > z t;\\\verb;u b > d t;\\\verb;u u~ > t~ t;\\\verb;c b > s t;\\\verb;c c~ > t~ t;\\\verb;d d~ > t~ t;\\\verb;s s~ > t~ t;\\\verb;b b~ > t~ t;\\\verb;b a > w- t;\\\verb;b ve > e- t;\\\verb;b vm > mu- t;\\\verb;b vt > ta- t;\\\verb;b z > w- t;\\\verb;a a > t~ t;\\\verb;a t > z t;\\\verb;ve ve~ > t~ t;\\\verb;vm vm~ > t~ t;\\\verb;vt vt~ > t~ t;\\\verb;e- e+ > t~ t;\\\verb;mu- mu+ > t~ t;\\\verb;ta- ta+ > t~ t;\\\verb;t t > t t;\\\verb;t z > z t;\\\verb;t w+ > w+ t;
\end{tabular}
\begin{tabular}[t]{l}
\hfil two-to-three\\\verb;g g  > t t~ h;\\\verb;u u~ > u u~ g;\\\verb;u u~ > u u~ a;\\\verb;u u~ > u u~ z;\\\verb;u u~ > u u~ h;\\\verb;u u~ > u d~ w-;\\\verb;u u~ > c c~ g;\\\verb;u u~ > c c~ a;\\\verb;u u~ > c c~ z;\\\verb;u u~ > c c~ h;\\\verb;u u~ > c s~ w-;\\\verb;u u~ > d u~ w+;\\\verb;u u~ > d d~ g;\\\verb;u u~ > d d~ a;\\\verb;u u~ > d d~ z;\\\verb;u u~ > d d~ h;\\\verb;u u~ > s c~ w+;\\\verb;u u~ > s s~ g;\\\verb;u u~ > s s~ a;\\\verb;u u~ > s s~ z;\\\verb;u u~ > s s~ h;\\\verb;u u~ > b b~ g;\\\verb;u u~ > b b~ a;\\\verb;u u~ > b b~ z;\\\verb;u u~ > b b~ h;\\\verb;u u~ > b t~ w+;\\\verb;u u~ > t b~ w-;\\\verb;u u~ > t t~ g;\\\verb;u u~ > t t~ a;\\\verb;u u~ > t t~ z;\\\verb;u u~ > t t~ h;\\\verb;u c~ > d c~ w+;\\\verb;u c~ > d s~ g;\\\verb;u c~ > d s~ a;\\\verb;u c~ > d s~ z;\\\verb;u c~ > d s~ h;\\\verb;u d~ > d d~ w+;\\\verb;u d~ > s s~ w+;\\\verb;u d~ > b b~ w+;\\\verb;u d~ > t b~ g;\\\verb;u d~ > t b~ a;\\\verb;u d~ > t b~ z;\\\verb;u d~ > t b~ h;\\\verb;u d~ > t t~ w+;
\end{tabular}
\begin{tabular}[t]{l}
\verb;c u~ > c d~ w-;\\\verb;c u~ > s d~ g;\\\verb;c u~ > s d~ a;\\\verb;c u~ > s d~ z;\\\verb;c u~ > s d~ h;\\\verb;c c~ > c c~ g;\\\verb;c c~ > c c~ a;\\\verb;c c~ > c c~ z;\\\verb;c c~ > c c~ h;\\\verb;c c~ > c s~ w-;\\\verb;c c~ > d d~ g;\\\verb;c c~ > d d~ a;\\\verb;c c~ > d d~ z;\\\verb;c c~ > d d~ h;\\\verb;c c~ > s c~ w+;\\\verb;c c~ > s s~ g;\\\verb;c c~ > s s~ a;\\\verb;c c~ > s s~ z;\\\verb;c c~ > s s~ h;\\\verb;c c~ > b b~ g;\\\verb;c c~ > b b~ a;\\\verb;c c~ > b b~ z;\\\verb;c c~ > b b~ h;\\\verb;c c~ > b t~ w+;\\\verb;c c~ > t b~ w-;\\\verb;c c~ > t t~ g;\\\verb;c c~ > t t~ a;\\\verb;c c~ > t t~ z;\\\verb;c c~ > t t~ h;\\\verb;c d~ > s d~ w+;\\\verb;c s~ > s s~ w+;\\\verb;c s~ > b b~ w+;\\\verb;c s~ > t b~ g;\\\verb;c s~ > t b~ a;\\\verb;c s~ > t b~ z;\\\verb;c s~ > t b~ h;\\\verb;c s~ > t t~ w+;\\\verb;d u~ > d d~ w-;\\\verb;d u~ > s s~ w-;\\\verb;d u~ > b b~ w-;\\\verb;d u~ > b t~ g;\\\verb;d u~ > b t~ a;\\\verb;d u~ > b t~ z;\\\verb;d u~ > b t~ h;\\\verb;d u~ > t t~ w-;
\end{tabular}
\begin{tabular}[t]{l}
\verb;d c~ > d s~ w-;\\\verb;d d~ > d d~ g;\\\verb;d d~ > d d~ a;\\\verb;d d~ > d d~ z;\\\verb;d d~ > d d~ h;\\\verb;d d~ > s s~ g;\\\verb;d d~ > s s~ a;\\\verb;d d~ > s s~ z;\\\verb;d d~ > s s~ h;\\\verb;d d~ > b b~ g;\\\verb;d d~ > b b~ a;\\\verb;d d~ > b b~ z;\\\verb;d d~ > b b~ h;\\\verb;d d~ > b t~ w+;\\\verb;d d~ > t b~ w-;\\\verb;d d~ > t t~ g;\\\verb;d d~ > t t~ a;\\\verb;d d~ > t t~ z;\\\verb;d d~ > t t~ h;\\\verb;s c~ > s s~ w-;\\\verb;s c~ > b b~ w-;\\\verb;s c~ > b t~ g;\\\verb;s c~ > b t~ a;\\\verb;s c~ > b t~ z;\\\verb;s c~ > b t~ h;\\\verb;s c~ > t t~ w-;\\\verb;s s~ > s s~ g;\\\verb;s s~ > s s~ a;\\\verb;s s~ > s s~ z;\\\verb;s s~ > s s~ h;\\\verb;s s~ > b b~ g;\\\verb;s s~ > b b~ a;\\\verb;s s~ > b b~ z;\\\verb;s s~ > b b~ h;\\\verb;s s~ > b t~ w+;\\\verb;s s~ > t b~ w-;\\\verb;s s~ > t t~ g;\\\verb;s s~ > t t~ a;\\\verb;s s~ > t t~ z;\\\verb;s s~ > t t~ h;\\\verb;b b~ > b b~ g;\\\verb;b b~ > b b~ a;\\\verb;b b~ > b b~ z;\\\verb;b b~ > b b~ h;\\\verb;b b~ > b t~ w+;
\end{tabular}
\begin{tabular}[t]{l}
\verb;b b~ > t b~ w-;\\\verb;b b~ > t t~ g;\\\verb;b b~ > t t~ a;\\\verb;b b~ > t t~ z;\\\verb;b b~ > t t~ h;\\\verb;b t~ > t t~ w-;\\\verb;t b~ > t t~ w+;\\\verb;t t~ > t t~ g;\\\verb;t t~ > t t~ a;\\\verb;t t~ > t t~ z;\\\verb;t t~ > t t~ h;
\end{tabular}
\end{scriptsize}
\caption{Illustrative sets of two-to-two SM processes involving at least a Higgs or a top-quark which can be used as basis for a comparison. Examples of two-to-three processes including $gg\to t\bar{t}h$ and $q\bar q\to q\bar q\; g/\gamma/Z/W/h$ ones.}
\label{tab:processes}
\end{table}

\section{Extension to the one-loop level}
\label{sec:loop}
Relying on the BLHA~\cite{Binoth:2010xt, Alioli:2013nda} conventions for loop amplitude providers, one can also extend the comparison of SMEFT amplitudes to the loop level.

A setup block denoted as \texttt{<weight\_blha>} is added to the \texttt{<weight>} header. In addition to BLHA2 parameters the value of the renormalization scale \verb;mu; is also specified. Following, for instance, the example provided in figure\,6 of \rcite{Alioli:2013nda}, one would obtain a format like that of \autoref{tab:blha}.

Colour- and helicity-summed pole residues would then be specified in a \texttt{(PoleCoeff0,\linebreak PoleCoeff1, PoleCoeff2)} triplet stored in \texttt{<wgt>} fields for each implementation, process, phase-space point, parameter point, and EFT order.
The squared terms correspond here to loop amplitudes with one EFT operator insertion squared against tree amplitudes featuring also one EFT operator insertion.

As mentioned in \autoref{item:squared-amplitudes}, one could consider comparing squared helicity amplitudes in the future.

\begin{table}
\begin{scriptsize}\begin{verbatim}
<LesHouchesEvents version="3.0">
  <header>
    <initrwgt>
      <weightgroup name='...' weight_name_strategy='...'>
        <weight id='...'>
          ...
          <weight_blha>
            InterfaceVersion        BLHA2
            MatrixElementSquareType CHsummed
            CorrectionType          QCD
            IRregularisation        DREG
            WidthScheme             ComplexMass
            EWScheme                alphaGF
            AccuracyTarget          0.0001
            CouplingPower QCD       2
            CouplingPower QED       0
            mu                      91.188         # renormalization scale
          </weight_blha>
        </weight>
        ...
      </weightgroup>
    </initrwgt>
  </header>
  ...
  <event>
     4      2 +5.7925728e-01 9.11180000e+01 0.00000000e+00 1.18400000e-01
           21 -1    0    0    0    0 +0.00e+00 +0.00e+00 +5.00e+02 5.00e+02 0.00e+00 0.00e+00 9.00e+00
           22 -1    0    0    0    0 +0.00e+00 +0.00e+00 -5.00e+02 5.00e+02 0.00e+00 0.00e+00 9.00e+00
           -6  1    1    2    0    0 +1.66e+02 -3.90e+02 +2.01e+02 5.00e+02 1.72e+02 0.00e+00 9.00e+00
            6  1    1    2    0    0 -1.66e+02 +3.90e+02 -2.01e+02 5.00e+02 1.72e+02 0.00e+00 9.00e+00
    <rwgt>
    <wgt id='mod1-sm-sq'> (+2.1166989e-01, +3.6769585e-01, -3.2187585e-01) </wgt>
    <wgt id='mod1-ctW-int'> (+3.6769585e-01, +1.5438777e+00, +2.1166989e-01) </wgt>
    <wgt id='mod1-ctW-sq'> (+1.5438777e+00, -3.2187585e-01, +1.5438777e+00) </wgt>
    ...
    </rwgt>
  </event>
  ...
</LesHouchesEvents>
\end{verbatim}\end{scriptsize}%
\caption{Extension of the LHE formatting for one-loop processes. A \texttt{<weight\_blha>} sub-header is added with BLHA information and the residues of $\epsilon$ poles are quoted in \texttt{<wgt>} fields as \texttt{(PoleCoeff0, PoleCoeff1, PoleCoeff2)} triplets.}
\label{tab:blha}
\end{table}

\clearpage
\section{Comparisons performed}
\label{sec:comparisons}
A list of the comparisons performed between different implementations, according to the guidelines of this note, is provided here with references to their detailed discussion.
It shall be updated as new comparisons become available.
\begin{itemize}

\item A comparison of top-quark interactions in \textsc{dim6top}, \textsc{SMEFTsim} and \textsc{SMEFT@NLO} UFO models obtained with a dedicated \textsc{MadGraph} plugin are provided at this \href{https://bazaar.launchpad.net/~rwgtdim6/mg5amcnlo/plugin_eft_contrib/revision}{url}.
More explanations are provided in this \href{https://bazaar.launchpad.net/~rwgtdim6/mg5amcnlo/plugin_eft_contrib/view/head:/example/compare_models_top.py}{file}.

\item The \textsc{Recola2} model file including the dimension-six operators in the Warsaw basis has been
cross-checked against the \textsc{SMEFTsim\_A\_U35\_MwScheme\_UFO\_v2\_1 UFO} model file for the
subset of operators that only involve gauge bosons and/or scalars. The numerical results
corresponding to the former model have been obtained with the \textsc{Recola2} matrix-element
provider, while \textsc{Madgraph5\_aMC@NLO} was used for the latter model. A sample output file
in the LHE format described above can be found at the
\href{https://recola.hepforge.org/modelfiles/smeftcmp.html}{url} for a few selected
processes. As can be seen from the LHE file, the only differences between the predictions
of the two models are some interference matrix-elements that however correspond to
numerical zeros.

\item A comparison of HEL and BSMC implementations is ongoing.

\end{itemize}


\theendnotes

\let\href\oldhref
\bibliographystyle{apsrev4-1_title}
\raggedright\bibliography{eft_validation_principles.bib}
\end{document}